\documentclass[12pt]{article}
\usepackage{graphicx}

\newcommand\pubnumber{SNSN-323-63}
\newcommand\pubdate{\today}

\def\pdinfn{INFN - Padova,\\
I-35131 Padova, ITALY}

\def\Title#1{\begin{center} {\Large #1 } \end{center}}
\def\Author#1{\begin{center}{ \sc #1} \end{center}}
\def\Address#1{\begin{center}{ \it #1} \end{center}}

\newcommand\pubblock{\rightline{\begin{tabular}{l} \pubnumber\\
         \pubdate  \end{tabular}}}
\newenvironment{Abstract}{\begin{quotation}  }{\end{quotation}}
\newenvironment{Presented}{\begin{quotation} \begin{center} 
             PRESENTED AT\end{center}\bigskip 
      \begin{center}\begin{large}}{\end{large}\end{center} \end{quotation}}
\def\Acknowledgements{\bigskip  \bigskip \begin{center} \begin{large}
             \bf ACKNOWLEDGEMENTS \end{large}\end{center}}

\def\beq{\begin{equation}}
\def\eeq#1{\label{#1}\end{equation}}
\def\eeqn{\end{equation}}

\def\beqa{\begin{eqnarray}}
\def\eeqa#1{\label{#1}\end{eqnarray}}
\def\eeqan{\end{eqnarray}}

\let\bar=\overbar

\def\Dslash{\not{\hbox{\kern-4pt $D$}}}
\def\dslash{\not{\hbox{\kern-2pt $\del$}}}

\def\msb{{\bar{\ssstyle M \kern -1pt S}}}

\begin{document}
\begin{titlepage}
\pubblock

\vfill
\Title{THE NEUTRINO MASS HIERARCHY FROM OSCILLATION}
\vfill
\Author{Luca Stanco
}
\Address{\pdinfn}
\vfill
\begin{Abstract}
The ordering of the neutrino mass eigenstates, also addressed as Mass Hierarchy (MH), is one of the most relevant issues
    in neutrino physics, currently under investigation by many proposals and experiments. In this short note focus will 
    be given to the different ways to determine MH from neutrino oscillation data in the near future. 
    A pragmatic strategy is suggested and two recent new methods of analysis are recalled. Statistical issues and concerns are also addressed,
    envisaging the necessity of more accurate studies and analyses.
\end{Abstract}
\vfill
\begin{Presented}
NuPhys2017, Prospects in Neutrino Physics\\
Barbican Centre, London, UK,  December 20--22, 2017
\end{Presented}
\vfill
\end{titlepage}
\def\thefootnote{\fnsymbol{footnote}}
\setcounter{footnote}{0}

\section{Introduction}

The ordering of the neutrino mass eigenstates is one of the most relevant issues, currently under investigation by many 
proposals and experiments. In the standard scenario and a widely usual convention the three neutrinos $\nu_1$, $\nu_2$ and $\nu_3$ are known to have relative masses measured as $\delta m^2_{21}= m^2_2 - m^2_1$ (historically named ``solar'' mass term) and $|\Delta m^2_{31}|= |m^2_3- m^2_1|\sim |m^2_3- m^2_2|$ 
(called ``atmospheric'' mass term). The sign of $\Delta m^2_{31}$ has not been measured yet, and that allows two different configurations for 
the mass eigenstates: either $m_1< m_2< m_3$ or $m_3< m_1< m_2$. That corresponds to have either one or two higher mass 
states, with huge consequences on the neutrino models~\cite{lucas,pdg}.  The mass ordering is usually identified as normal 
hierarchy (NH) when $\Delta m_{31}^2 > 0$ and inverted hierarchy (IH) for the case $\Delta m_{23}^2 > 0$. 
Its importance is enormous to provide inputs for the next studies and
experimental proposals, to finally clarify the needs and the tuning of new projects, and to constraint analyses in other fields
like cosmology and astrophysics.

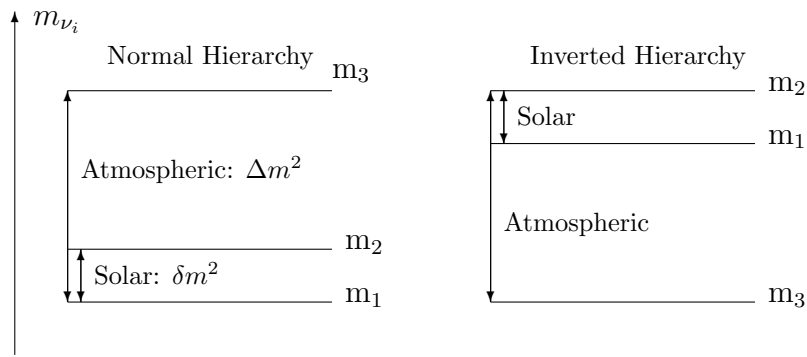
\begin{figure}[h]\label{fig1}
\centering
\begin{picture}(300,150)
\put(0,0){\vector(0,1){130}}\put(7,125){$m_{\nu_i}$}
\put(20,20){\line(1,0){100}}\put(20,40){\line(1,0){100}}\put(20,100){\line(1,0){100}}
\put(125,20){m$_1$}\put(125,40){m$_2$}\put(120,105){m$_3$}
\put(25,20){\vector(0,1){20}}\put(25,40){\vector(0,-1){20}}\put(20,20){\vector(0,1){80}}\put(20,100){\vector(0,-1){80}}
\put(35,110){\footnotesize Normal Hierarchy}\put(30,27){\footnotesize Solar: $\delta m^2$}\put(25,67){\footnotesize Atmospheric: $\Delta m^2$}
\put(180,20){\line(1,0){100}}\put(180,80){\line(1,0){100}}\put(180,100){\line(1,0){100}}
\put(285,20){m$_3$}\put(285,80){m$_1$}\put(285,100){m$_2$}
\put(180,20){\vector(0,1){80}}\put(180,100){\vector(0,-1){80}}\put(185,80){\vector(0,1){20}}\put(185,100){\vector(0,-1){20}}
\put(195,110){\footnotesize Inverted Hierarchy}\put(185,47){\footnotesize Atmospheric}\put(190,87){\footnotesize Solar}
\end{picture}
    \caption{Neutrino mass eigenstates for normal and inverted mass ordering (not to scale).}
\end{figure}

In Fig.~\ref{fig1} a cartoon of the two possible configurations for the mass ordering is depicted. In this paper
the following notation has been used for the atmospheric mass: $\Delta m^2_{atm}=\Delta m^2_{31} ({\rm NH})=
\Delta m^2_{23} ({\rm IH})$, for the two different hypotheses, respectively.
$\Delta m^2_{atm}$ is therefore a fundamental physical quantity, which corresponds to the difference of the heaviest squared
mass and the lightest neutrino squared-mass.

The achievements of the last two decades brought up a coherent picture, namely the oscillation of three neutrino
flavour--states, $\nu_e$, $\nu_{\mu}$ and $\nu_{\tau}$, originated by the mixing of the three  $\nu_1$, $\nu_2$ and $\nu_3$ 
mass eigenstates. 
The issue of the mass ordering has been highly debated in the last decade, but it gained in interest
with the discovery of the relatively large value of $\theta_{13}$ in 2012. The convolutions between the three mixing angles and 
the mass parameters are such that  measurements of the current experiments may become sensitive to the 
dependences of the oscillation probabilities to the sign of MH. 
Surely, the MH determination will be a major issue for the next experiments under construction.

\section{The MH degeneracies}

As far as oscillations are concerned, the dependences on the mass ordering come from the interference between two different
effects. In particular, the interference of oscillations driven by $\Delta m^2_{31}({\rm NH})$ or $\Delta m^2_{23}({\rm IH})$
with oscillations driven by another quantity, $Q$, with a known sign.
In vacuum the interference is given by the joint atmospheric and solar oscillations, such that $Q$ corresponds to the
solar mass $\delta m^2$. For atmospheric and neutrinos from long baseline accelerator the interference is due
to the matter effect, $Q$ being the corresponding matter potential, $2\sqrt{2} G_F N_e E$, with obvious meaning for the quantities
involved.
Moreover, in the three--neutrino framework MH is highly correlated with the neutrino oscillation parameters 
and the CP phase, $\delta_{CP}$.
Specifically, in the neutrino oscillation framework there are three big area of investigation: MH from long baseline 
accelerators are highly coupled to $\delta_{CP}$, while for the reactor antineutrino (medium baseline) there is no
$\delta_{CP}$ dependence at first order, in contrast to a strong dependence on the exact value of $\Delta m^2_{atm}$. 
The third area of
investigation corresponds to the atmospheric neutrinos, which own a degeneracy both on $\Delta m^2_{atm}$ and the 
value of the mixing angle $\theta_{23}$, namely to which octant it belongs.

\begin{figure}[h]\label{fig2}
\centering
\begin{center}
 \includegraphics[width=0.6\textwidth]{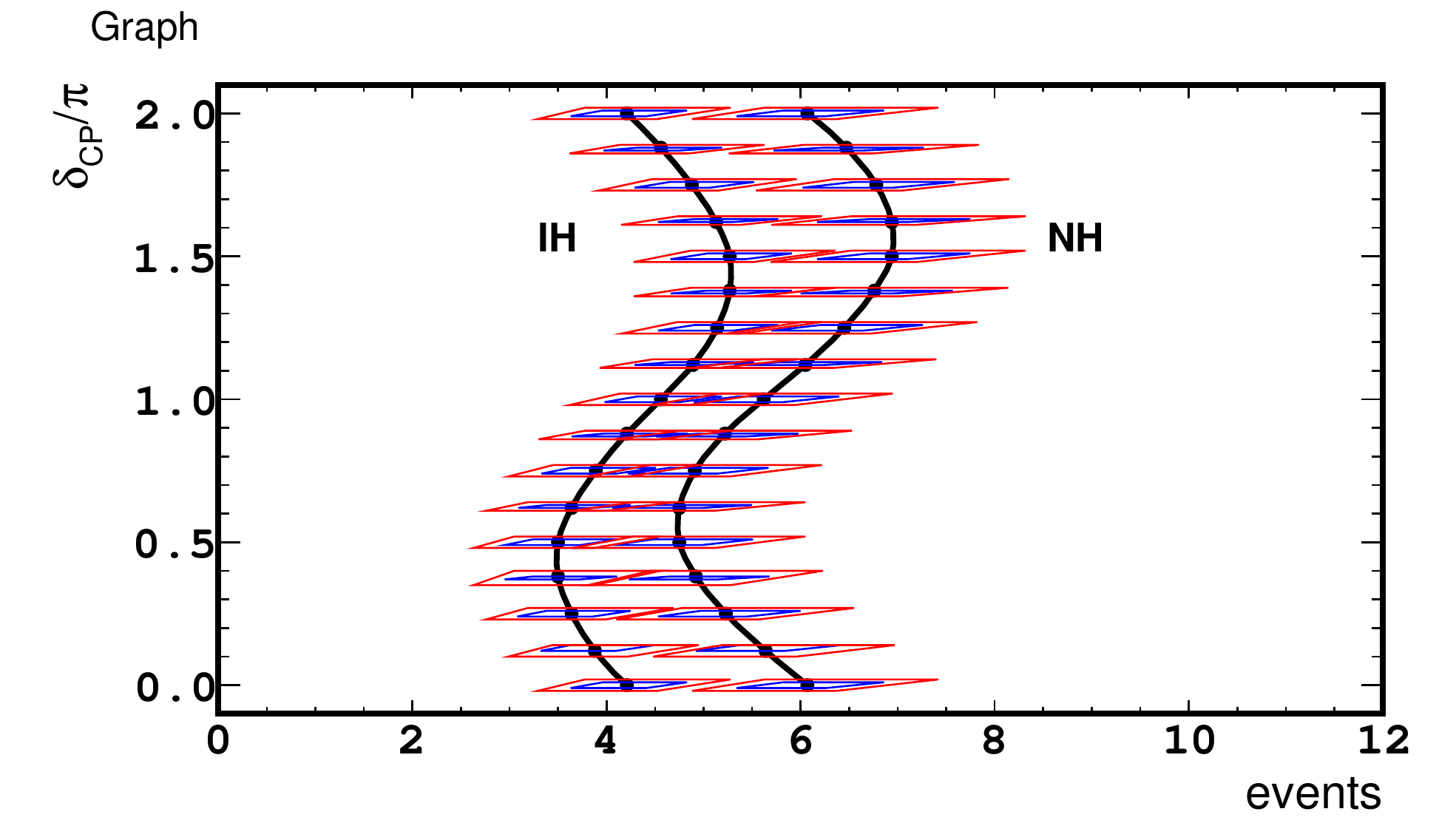}%
    \caption{The number of events ($\nu_{\mu}\rightarrow \nu_e$ appearance plus background) as function of $\delta_{CP}$ as expected for the
     2015 NOvA data analysis from the best fits of the 2017 global fit analysis (GF). The two plain black lines correspond
to the expectation from NH (right) and IH (left), respectively.
Our computation has been performed with the GLoBES package. The two concentric
areas for each value of $\delta_{CP}$ correspond to the 1 $\sigma$ and 2 $\sigma$ contours due to the correlated
$\theta_{23}$, $\theta_{13}$ uncertainties, $\sigma$ being estimated 
by GF. See~\cite{lucas1} for more details.}
\end{center}
\end{figure}

These correlations correspond to degeneracies that can severely limit the discrimination of the hierarchy, either normal of inverted. 
If one generally
indicates with $\theta_i$ the {\em correlation} parameter ($\theta_1=\delta_{CP}$, $\theta_2=\Delta m^2_{atm}$ and
$\theta_3=\theta_{23}$) more solutions may be extracted from the data for MH, 
e. g. NH($\hat{\theta_i}$) and IH($\hat{\theta'_i}$) with $\hat{\theta_i}\ne \hat{\theta'_i}$. The $\theta_i$ parameters are usually evaluated
within the standard $3\,\nu$ oscillation framework via global fits~\cite{gf}. Unfortunately, the current uncertainties on $\theta_i$ 
allow several distinct solutions and practically no sensitivity to MH. This is demonstrated in Fig.~\ref{fig2} for the NOvA case
and its 2015 data release.

\section{A strategy for the MH determination}

Given the scenario described in the previous section it is fundamental to control the test statistic that is used in the analysis. 
The statistical estimator should be robust and should make evidence of the degeneracies and the $\theta_i$ dependences. 
For the MH studies only one estimator has been extensively used so far throughout the several fields of investigation. That is
the chi-square difference,
$\Delta\chi^2= \chi^2_{min}({\rm IH})-\chi^2_{min}({\rm NH})$,
where the two minima are evaluated spanning the uncertainties of the three-neutrino oscillation parameters,
namely the solar mass $\delta m_{21}^2$, the atmospheric mass $\Delta m_{31(23)}^2$,
the CP phase $\delta_{CP}$ and  the mixing angles $\theta_{12}$, $\theta_{23}$,
 $\theta_{13}$, as defined by the standard parameterization. On top of that statistical and systematic errors are included
in the fitting procedures.
The $\Delta\chi^2$ evaluation is based on two distinct hypotheses, NH and IH. For each MH  the best solution is found: 
the  $\chi^2_{min}$ comes from two different best-fit values for NH and IH, separately, and the $\Delta\chi^2$ is the result 
of the internal
adjustments of the two distinct fits. No real understanding of the weight arising from each single contributions 
(i.e. the single neutrino oscillation parameters or the statistical/systematic errors) is possible, given to the intrinsic 
multiple non-linear correlations.
 
Recently, we suggested a change of perspective: try to identify an estimator that couples NH/IH and decouples the 
$\theta_i$ dependences~\cite{lucas}. As a consequence, each kind of data, long baseline or reactors or atmospheric ones,
should be analyzed by different optimized estimators. We already studied possible new estimators for the accelerator 
data~\cite{lucas1} and for reactor antineutrinos~\cite{lucas2}. Since these estimators already intrinsically couple
NH and IH, it is no more necessary to construct an ``estimator of the estimators'' like the $\Delta\chi^2$. Instead, the two 
outcomes, one for NH and the other for IH, are directly used to get NH and IH significances (using event-by-event Monte Carlo simulation to determine their probability distributions).

The change of perspective suggests a pragmatic new strategy in the determination of MH. Once the statistical estimator 
has been chosen, let us call $S$, its evaluation over data would simply bring to one of the three following options: 
\begin{enumerate}
\item both $S_{\rm NH}$ and $S_{\rm IH}$ are  {\em compatible} with data;
\item both $S_{\rm NH}$ and $S_{\rm IH}$ are  {\em incompatible} with data;
\item either $S_{\rm NH}$ or $S_{\rm IH}$ is compatible with data, the other one being incompatible.
\end{enumerate} 
The meaning of {\em compatible} and {\em incompatible} comes from a long experience in data analysis of experiments 
in particle physics. 
Nowadays, it is well accepted that {\em compatible} means at 95\% of C.L., whereas {\em incompatible}
means $\ge 5\, \sigma$. That corresponds to the standard definition of exclusion or observational results~\cite{cowan}.
Over the last decades, these choices have been proven to be the right ones by many experimental results.
To be more precise, an experimental observation to be conclusive corresponds to the rejection of the background hypothesis
at least at $5\, \sigma$. An experimental exclusion limit corresponds to the phase space defined by the set of values 
of the signal parameter compatible at 95\% C.L. with the data themselves, the complementary phase space containing the rejection 
of the signal at 95\% C.L..

When this procedure is applied to the MH determination, a confusing scenario may rise up. The question becomes: the MH
determination is a signal or a background rejection? Since NH/IH are mutually exclusive not-nested hypotheses
their roles can be interchanged. Then, our proposal is just the above list of options. Specifically, a conclusive experiment, or a
global analysis, should provide both a rejection of the wrong hierarchy at $5\, \sigma$ level and a compatibility with the 
true hierarchy at 95\% C.L.. 

When the analysis should produce a result as in case (1), thus it would be inconclusive. In case (2) probably something wrong 
were occurring in the analysis procedure (or the 3$\nu$ framework is no more appropriate). Case (3) should correspond
to the sensitivity with which the experiment/analysis would determine the mass hierarchy. In case (3) and a sensitivity
at the level of $5\, \sigma$ the determination of the MH could be finally established.

We also outline that different statistical approaches may be applied, namely a frequentist approach
or a Bayesian one. Only when the significance reaches the level of $5\, \sigma$ the different statistical approaches
usually give similar results.

\section{An estimator for MH at accelerators}\label{sec:nova}

For the accelerator basis searches the NOvA experiment is the best placed one~\cite{nova-patt}.
It is foreseen that some information be available after several years of running with data-taking both in neutrino and anti-neutrino modes. 
Adding measurements on $\delta_{CP}$ from few years of T2K exposure 
will allow to slightly increase the separation between the two options in different portions of $\delta_{CP}$ range~\cite{nagaya}.
Conversely, if MH should be known sometimes in future, T2K would greatly improve its significance on 
$\delta_{CP}$~\cite{t2k-future}.
The expectation on MH is however not exciting; only a 3-sigma significance could be obtained and only in the most favorable
$\delta_{CP}$ regions.
As a matter of fact the perspectives in the near future for the determination of the neutrino mass ordering with neutrinos 
from accelerator beams are rather 
poor, even less favorable than the prospects for the $\delta_{CP}$ measurement.

The new technique reported in~\cite{lucas1} is based on a new test statistic that properly weights the intrinsic statistical 
fluctuations of the data and extracts the significances of either NH or IH.
The Poisson distributions for $n_i$ observed events, $f_{\rm MH}(n_i;\mu_{\rm MH} |\delta_{CP})$ are initially considered,
where $\mu_{\rm MH} (\delta_{CP} )$ are the expected number of events as function of $\delta_{CP}$,
 MH standing for IH or NH. 
For a specific $n$ the left and right cumulative functions of $f_{\rm IH}$ and $f_{\rm NH}$ are computed and their ratios, $q_{\rm MH}$, are evaluated. Since for the $\nu_e$ appearance at NOvA the 
number of expected events as function of $\delta_{CP}$ is asymmetric towards IH and NH (less events are expected for IH than for NH), the ratios are defined independently for the IH and the NH cases:

\vspace{-0.1cm}
 {\footnotesize
\[
q_{\rm IH}(n, \delta_{CP}) =\frac{\sum_{n_i^{\rm IH}\ge n} f_{\rm IH}(n_i^{\rm IH};\mu_{\rm IH} |\delta_{CP} )}{\sum_{n_i^{\rm NH}\ge n}
f_{\rm NH}(n_i^{\rm NH};\mu_{\rm NH} |\delta_{CP} )},
\]
\[
q_{\rm NH}(n, \delta_{CP}) = \frac{\sum_{n_i^{\rm NH}\le n} f_{\rm NH}(n_i^{\rm NH};\mu_{\rm NH} |\delta_{CP})}{\sum_{n_i^{\rm IH}
\le n} f_{\rm IH}(n_i^{\rm IH};\mu_{\rm IH} |\delta_{CP})}. 
\]
}
$q_{\rm IH}$ and $q_{\rm NH}$ are two discretized random  variables comprised to the [0, 1] interval. As $n$ goes to zero  
$q_{\rm IH}$
goes to one, while when $n$ increases $q_{\rm IH}$ asymptotically tends to zero. $q_{\rm NH}$ behaves the other way around 
towards $n$.

The probability mass functions, $P(n)$, of each $q_{\rm MH}$
have been computed via toy 
Monte Carlo simulations based either on $f_{\rm IH}$ (test of IH against NH) 
or $f_{\rm NH}$ (test of NH against IH).
They are further compared to the real number of observed data $n_D$. By evaluating the $p$--value probabilities for $n_D$
the significance is finally computed.

With the new method an averaged increase  of 0.5 $\sigma$ with respect to the standard $\Delta\chi^2_{min}$ is obtained~\cite{lucas1}.
Worth to note that the increase is not constant but it depends on the discrimination threshold $n_D$ and $\delta_{CP}$: 
the gain of the new method in terms of the number of sigma's strongly raises with $n_D$ and ``favorable''
regions of $\delta_{CP}$. As demonstrated in the appendix of~\cite{lucas1} the new method is generally better than $\Delta\chi^2_{min}$
for many reasons: it deals with the full probability distributions, it profits of the intrinsic fluctuations of the data and, most relevant,
it answers the right question (to disprove one MH option). In fact the new $q$ estimator focusses on the possibility to reject
the wrong hierarchy, disregarding the other one. Therefore, once one option is selected (e.g. rejection of IH) it does not provide any 
evaluation on the other option (rejection of NH). Instead, the $\Delta\chi^2_{min}$ method  treats the two options
in a symmetric way with the disadvantage of mixing up the information.

This new procedure becomes asymptotically equivalent to the $\Delta\chi^2$ one when the luminosity increases. Nevertheless,
it allows to achieve a similar level of significance with about a factor three less data.

\section{An estimator for MH at reactors}\label{sec:juno}

The determination of the mass hierarchy with reactor neutrinos is a very challenging task. Both an exceedingly high energy-resolution and
a large mass detectors are required. The JUNO experiment has been proposed and it is currently under construction 
trying to match these two requirements~\cite{juno}. The foreseen achievement on MH is nevertheless limited. A significance 
around $4\,\sigma$,
after 6 years of exposure, at the full reactor power of 36 GW, is predicted for the median sensitivity.

A new technique that would provide a robust 5 $\sigma$ measurement in less than six years of running was recently proposed 
in~\cite{lucas2}. It is based on the introduction of a new statistical estimator, F, which revises the approach followed in the
last ten years based on the $\Delta\chi^2$ estimator and the effective parameterization of the neutrino masses.
The effective parameterization~\cite{parke} was very valuable in boosting the studies and the proposals
for a large reactor neutrino experiment at medium baseline (30-50 km). It predicted the possible determination of MH
without any degeneracy, even taking into account the rather large uncertainty on $\Delta m^2_{atm}$ at that time
(larger than 40\%). However, the effective parameterization reduces the information above a certain neutrino energy threshold.
For example, at JUNO the discrimination between NH and IH vanishes for $E_{\nu}> 4 -5 MeV$.

Instead, using the F estimator the two mass orderings could be discriminated at the price of allowing for two different values of $\Delta m^2_{atm}$. 
This degeneracy on $\Delta m^2_{atm}$ (around $12\times 
10^{-5}$ eV$^2$) can in any case be measured at an unprecedented accuracy of much less than 1\%, i.e. $10^{-5}$ eV$^2$,
within the same analysis. 

The key picture is shown in Fig.~3. It demonstrates that, whatever be the algebraic construction and the implicit
assumptions of the F-test, for each real/simulated data sample (x-axis) the F technique identifies two main possibilities
(y-axis): the true MH associated to the true $\Delta m^2_{atm}$ and a wrong MH solution with a $\Delta m^2_{atm}$
shifted of $12\times 10^{-5}$ eV$^2$ with respect to the true value. Each of the two solutions own a $\Delta m^2_{atm}$
resolution of about 0.3\%.

\begin{figure}[h]\label{fig3}
\centering
\begin{center}
\includegraphics[width=7.5cm]{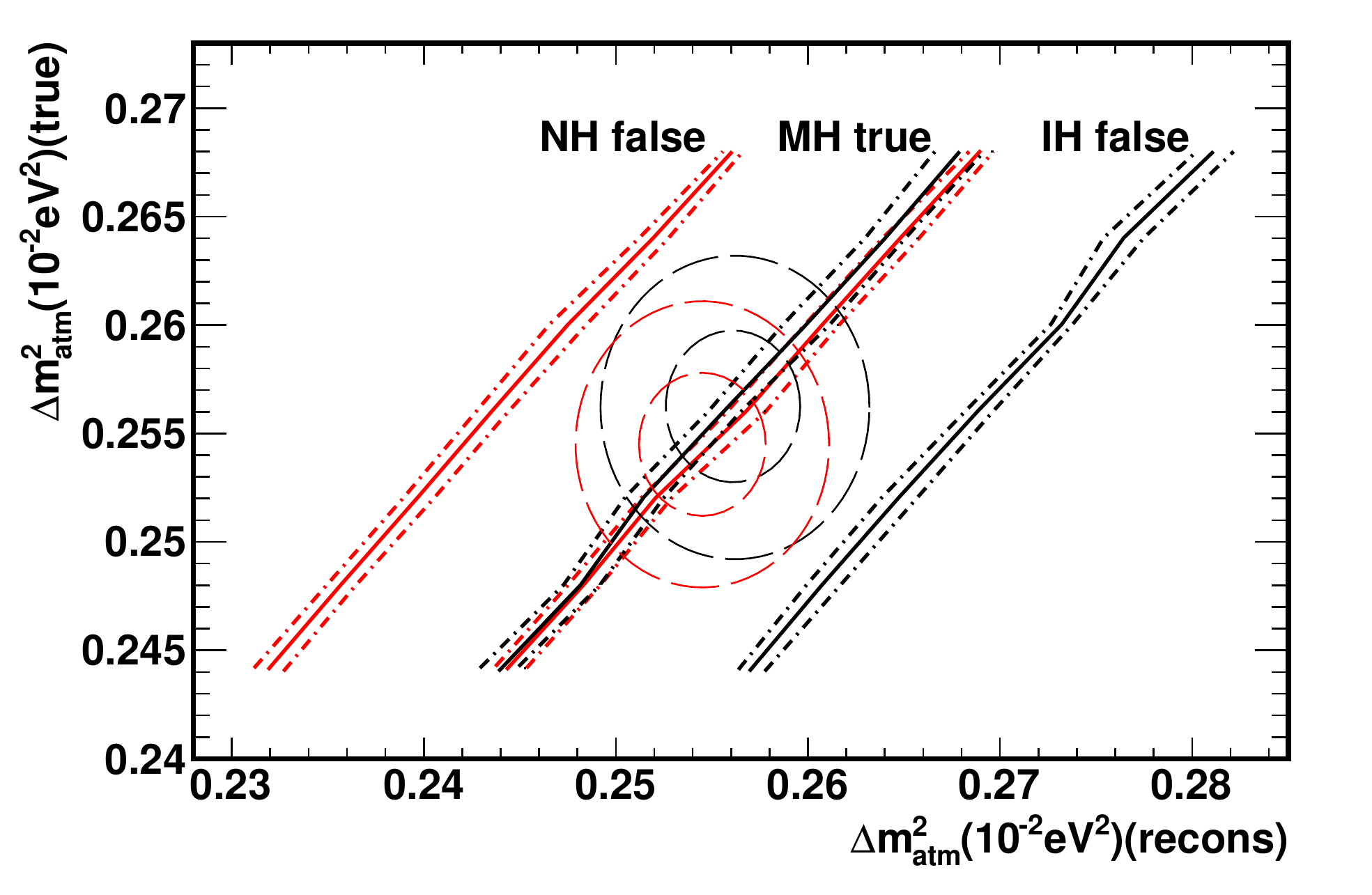}
\caption{$\Delta m^2_{atm}$(true) vs $\Delta m^2_{atm}$(recons) is drawn,
$\Delta m^2_{atm}$(recons) being obtained by applying the $F$ estimator. The continuous lines correspond to the
central values, the dashed ones to the $\pm\,\sigma$ bands. Black (red) curves corresponds to the NH (IH) generation.
The central circles correspond to the 68\% and 95\% C.L.
contours of the current $\Delta m^2_{atm}$ uncertainties from the global fits for NH and IH.
See~\cite{lucas2} for more details.}
\end{center}
\end{figure}

It is worth to add that in~\cite{lucas2} evaluation and inclusion of systematic errors and backgrounds have been performed, 
the most relevant among them 
being the addition of the two remote reactor plants 250 km away. Baselines of each contributing
reactor core and its spatial resolution have been taken into account. 
Possible results after two years of running and the foreseen initially-reduced available reactor power have been studied, too.
The Monte Carlo simulations have been performed following an event-by-event procedure.

Last but not least, using the F estimator, the estimated significance grows with the size of the data sample, contrary
to the $\Delta\chi^2$ outcome that is asymptotically limited.

\section{Two words on the MH sensitivity estimation}

The conventional way to establish the MH  sensitivity is to follow the frequentist prescription: the median discovery-significance 
expected from an experiment is computed as the $p$-value of the {\em background} probability density function (PDF)
corresponding to the median of the PDF of the {\em signal}, sometime with $\pm 1\, \sigma$ bands. That gives the 50\% probability 
that the experiment achieve such significance. The final significance may be higher or lower. 
This procedure is quite useful to compare different experimental proposals but it may be limited when e.g. more robust 
expectations are required for the optimizations of the experiment. That becomes more relevant when the significance level 
is critical, that is between 3 and 5 $\sigma$'s.

We prefer and suggest that for the MH determination one should assume a certain confidence level C for the true hypothesis
and an average $p$-value for the wrong hypothesis be evaluated, weighted by the true-hypothesis PDF. 
The single $p$-value entering in the average is computed from the edge of the confidence interval of the true-hypothesis PDF.
The new C parameter that enters in the computation is driven by the experimental confidence on the quality of the
experiment itself. C may be chosen to be 68\% or 90\%, depending of the risk approach, or even 99\% if concerns about
possible systematics have to be taken into account.
In formulas, being $f_{MO}(\vec{x})$ the density probabilities and considering the case NH true,
\begin{eqnarray}
&p-val(IH)=\int\limits_{\Omega_{NH}}\mathrm{d}\vec{x}\, f_{NH}(\vec{x}) \otimes \int\limits_{\Omega_{NH}(\vec{x})}
\mathrm{d}\vec{x}' \, f_{IH}(\vec{x}') , \nonumber \\
&\Omega_{NH}\perp  \int\limits_{\Omega_{NH}} \mathrm{d}\vec{x}\, f_{NH}(\vec{x}) = C, \quad
\Omega_{NH}(\vec{x})\ni  f_{IH}(\vec{x}')\le\, f_{IH}(\vec{x})\quad \mathrm{for}\quad \vec{x}'\in \Omega_{NH}. \nonumber 
\end{eqnarray}
\noindent This procedure is in general more conservative than the evaluation of the standard $p$-value on the median
of the true hypothesis. 

Another issue about the MH sensitivity determination regards the common studies with the $\Delta\chi^2$ estimator.
A big warning should be addressed to the figure of merits obtained from the analytical or semi-analytical developments.
All these analyses usually assume that the asymptotic distributions of the likelihood ratio test statistic
given in~\cite{cowan} are a valid approximation~\cite{read}. 
This is a generalization of the approximations derived in~\cite{appr1} for the reactor experiments to evaluate the distribution of 
the $\Delta\chi^2$, under the assumption that the data follow a Gaussian distribution in the large sample limit. 
In other words, the Central Limit Theorem (CLT) 
should hold. 

A strict condition of the CLT is that the variance of the PDF of the single measurement be constant\footnote{A counter example is the evaluation of the particle momentum $q$ from the measurement of the curvature of the track in space: the Gaussian
distributions of the spatial coordinate $x$ will never produce a Gaussian distribution for the PDF of $q$
as the variance of $q$ varies with $1/x^4$.}. 
Therefore,
the Gaussianity could be destroyed when systematics are included. In particular, when the systematic
errors are largely varying, like e.g. the case of Juno, the Gaussian approximation can be badly broken and reduce
notably the significances. The net result is not easily noticeable as the medians and the Asimov data
sets  are not affected. An event-by-event simulation that properly takes into account the convolution of statistical
fluctuations and systematic errors is needed to determinate the correct distributions~\cite{lucas3}.

\section{Conclusions}\label{sec:conc}

A major enterprise of the neutrino community is the future determination of the neutrino mass ordering.
Unfortunately, it appears to be a challenging task for any framework should be used. The atmospheric
neutrino framework, as well as the cosmological framework, were not discussed in this note. Nevertheless
their standard sensitivities on the MH determinations seem either rather poor or with severe concerns, respectively.
The same occurs to the framework of the accelerator baseline and the reactor neutrinos.
Therefore, it is mandatory to evaluate whether new tools of analysis can overcome these limits.
We reported about the recent techniques developed for the two latter frameworks. They are encouraging
and could provide more robust and significant/complementary results than the standard technique based on the
$\Delta\chi^2$ estimator, and in a shorter time.

\Acknowledgements
It is a pleasure to thank the organizers for their kind invitation,
the warm hospitality in London, and the overall very stimulating series of presentations
at this NuPhys2017 conference.
I would like to acknowledge discussions about the issues illustrated in this brief note
with my colleagues S. Dusini, A. Lokhov, G. Salamanna C. Sirignano and M. Tenti. The help of F. Sawy in
some lateral work about the reactor neutrino study is also acknowledged, as well as deep
discussions with S. Petcov.

\end{document}